\documentstyle[prb,aps,twocolumn,amssymb,floats]{revtex}

\begin{document}

\title {\bf Metal-insulator transition at B=0 in p-SiGe}
\author{
       P.T. Coleridge, R.L. Williams, Y. Feng and P. Zawadzki\\ }

\address
{Institute for Microstructural Sciences, National Research Council, Ottawa,
Ontario, K1A OR6, Canada\\ }

\date{31 July 1997}

\maketitle

\begin{abstract}
{\bf
Observations are reported of a metal-insulator transition in a 
two-dimensional hole gas in asymmetrically doped strained SiGe quantum wells.
The metallic phase, which appears at low temperatures in these 
high mobility samples, is characterised by a resistivity that decreases
exponentially with decreasing temperature. This behaviour, and the duality
between resistivity and conductivity on the two sides of the transition,
 are very similar to that recently reported for high mobility Si-MOSFETs.
}
\end{abstract}

      In high mobility Si-MOSFETs scaling measurements,
as a function of both temperature and electric field\cite{r1,r2} have shown
that there is a well  defined metal-insulator
transition at B = 0  with a critical resistivity of order
h/e$^2$. The transition is interesting not only because it
contradicts the commonly accepted view that scaling theory\cite{r3}
predicts all states in a disordered noninteracting two-
dimensional system to be localised at B = 0 but also
because on the metallic side of the transition the
resistance is unconventional in the sense that it
decreases exponentially with decreasing temperature. This
strong enhancement of the conductivity closely parallels
the exponential increase in resistivity on the insulating
side of the transition and strongly suggests they have a
common origin\cite{r2}. The behaviour appears only in samples
with a high mobility\cite{r4} above about 1 m$^2$/Vs. A recent
experiment\cite{r5} has shown this explicitly by independently
varying the sample mobility using a substrate bias.

       Two recent developments have pointed to a
resolution of the apparent conflict between these results
and one-parameter scaling theory. For interacting
electrons a recent paper\cite{r6} shows that a 2D metal-insulator
transition does not contradict general scaling principles
but that the metal is likely to not be a Fermi liquid.
Secondly, it has been argued that at low temperatures
scattering in Si-MOSFETs is dominated by a "spin-gap"
associated with strong spin-orbit coupling\cite{r4}. In this
case, where a sympletic ensemble is involved, rather than
unitary or orthogonal, the scaling function $\beta(g)$ (given by
$d[\ln(g)]/d[\ln(L)]$ where g is a conductance and L the size
of the system) remains positive in the large g (low
disorder) limit\cite{r7}. As the disorder increases and $\beta$
eventually becomes negative a metal-insulator transition
can occur. The "spin-gap" is well developed when large
angle scattering dominates and a significant fraction of
the scattering events involve a reversal of the k-vector
and an effective spin-flip.
        
       Evidence for a metal-insulator transition at B = 0
has also been seen in p-type modulation doped strained
SiGe quantum wells. In symmetrically doped wells this has
been observed both as a function of well width\cite{r8} and, in a
gated sample, as a function of density\cite{r9}.  Results are
reported here of transport measurements in asymmetrically
doped wells that show the same kind of transition with a
resistivity in the metallic phase that decreases
exponentially with temperature. In both Si-MOSFETs and p-
SiGe the energy associated with many-body interactions is
large compared with the kinetic energy of the carriers:
for example, in p-SiGe, at a density of 1$\times$10$^{15}$ m$^{-2}$
the hole-hole interaction energy is about 6.5 meV while the
Fermi energy is only 0.5meV. Also in p-SiGe the holes are
in almost pure $|M_J|$ = 3/2 states so as in the Si-MOSFETS
spin-orbit effects are likely to play a significant role.

\begin{table*}[tbh]
\caption
{Sample parameters. The conventional (transport) mobility $\mu_{tr}$ is the
maximum value measured. The quantum mobility $\mu_q$ is derived from the low
field SdH oscillations measured at approximately 50mK.}

\begin{tabular}{cccdddd}

Sample  &Growth         &Spacer         &Hole Density
        &$\mu_{tr}$(max)  &$\mu_q$        &m*/m$_0$       \\
        &               &(nm)           &(10$^{15}$m$^{-2}$)
        &(m$^2$/Vs) &(m$^2$/Vs)     \\
                                                        \hline
A       &CVD121     &20     &3.05    &1.13      &1.42      &.28--.32 \\
B       &CVD191     &12     &2.8      &1.51     &1.4       &.29  \\
C       &CVD193     &20     &1.5      &1.87     &1.41      &.24  \\
D       &CVD192     &36     &1.2      &1.18     &1.30      &.22  \\
E       &CVD276     &20     &1.3      &$>$1.18  &-      &-  \\
F       &CVD275     &20     &0.48\tablenotemark[1] &0.99     &-      &-  
                                                        
\end{tabular}
\tablenotetext[1]{Depleted in the dark, density after a small amount of
illumination}
\label{table1}
\end{table*}

       The samples, grown in a UHV-CVD system, consisted
of a n- substrate with a 300nm Si buffer layer and a 40nm
Si$_{.88}$Ge$_{.12}$ quantum well. A spacer layer on top of the well
was followed by a 30nm  Si(B) layer with doping that
varied between 0.5 and 3$\times$10$^{24}$m$^{-3}$. The SiGe well is
compressively strained and the asymmetric doping means the
holes reside in an approximately triangular potential
well. Measurements were made on Hall Bar samples (width
200 $\mu$m) with Al contacts alloyed in at a temperature of
approximately 540$^0$C. Sample parameters are listed in
table \ref{table1}. Low temperature illumination using a red LED
usually produced a small persistent photoconductivity (PPC) effect
which could be exploited to produce small increases in the
density.  The measured hole densities are all larger than
can be explained in terms of the standard electrostatic
model\cite{r10} using a valence band offset of 120meV and an
acceptor binding energy of 30meV (appropriate for B in
Si). Also the mobilities are smaller than theoretically
expected for scattering dominated by the remote ionised
acceptors\cite{r11}. The results are, however, consistent with
additional negatively charged impurities, of unknown
origin, at the interface as has been previously 
suggested\cite{r11,r12}. Quantum mobilities, determined from the low field
Shubnikov-de Haas (SdH) oscillations\cite{r13}, are all within
about 20\% of the peak transport mobilities which is
consistent with large angle scattering produced
predominantly by these impurities rather than by the
remote ionised acceptors. Effective mass values, deduced
from the temperature dependence of the low field SdH
oscillations, are in the range 0.2 -- 0.3m$_0$ although there
is a small uncertainty ($\sim$ 10\%) in the interpretation of
these measurements associated with the temperature
dependence of the B = 0 resistivities. 

       In all samples, the low field SdH oscillations are
dominated by minima that occur at odd filling factors.
This is a well known phenomenon in this system\cite{r14} caused
by a spin-splitting that is larger than half the cyclotron
spacing. Measurements in tilted fields\cite{r15,r16} show
that at low fields the cyclotron splitting and the spin-splitting
both depend only on the perpendicular component of
magnetic field. This confirms that strain and confinement
have raised the heavy-hole, light-hole degeneracy at the
zone centre so that the holes lie in a split-off heavy-
hole band (J = 3/2, $|M_J| = 3/2$) with very little
admixture from other bands.

\begin{figure}[tbh]
\vspace{8.5cm}
\includegraphics{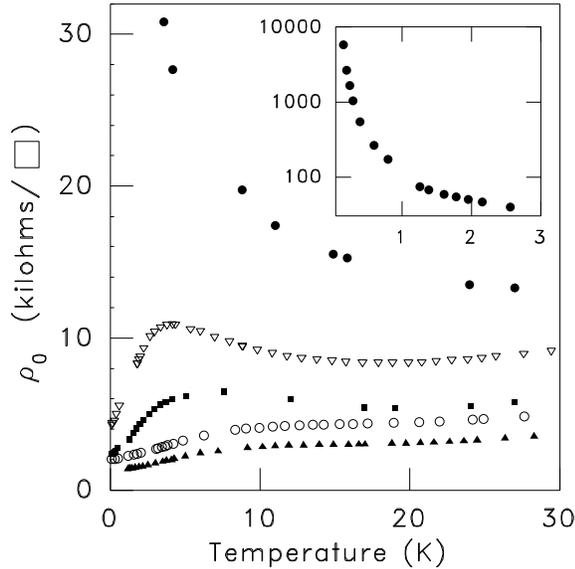}
\caption {
Temperature dependence of the zero field resistivity for samples
A($\circ$);B($\blacktriangle$);C($\blacksquare$); D($\triangledown$) and
F($\bullet$). Inset shows low temperature data for sample F with a
logarithmic scale.
}
\label{fig1}
\end{figure}

       Figure~\ref{fig1} shows the temperature dependence of the
zero field resistivity. At high densities the behaviour is
dominated by a monotonic decrease with temperature
associated with a reduction in the impurity scattering
rate\cite{r17}. There is also a peak around  5 -- 10K which
systematically develops as the density is reduced. A
similar peak is seen in Refs. 9 and 12 although in the
latter case it was attributed to changes in screening. The
peak results from  competition between insulating
behaviour ($d\rho/dT < 0$), which develops as the density is
reduced, and a new type of strongly enhanced conductivity
that dominates at lower temperatures (below about 4K). The
enhanced conductivity produces a drop in resistance,
typically by a factor of about 3, and metallic behaviour
($d\rho/dT > 0$). In the lowest density sample, however, the
insulating behaviour dominates and $d\rho/dT$ remains negative
at least down to 0.1K.

\begin{figure}[tbh]
\vspace{6.7cm}
\includegraphics{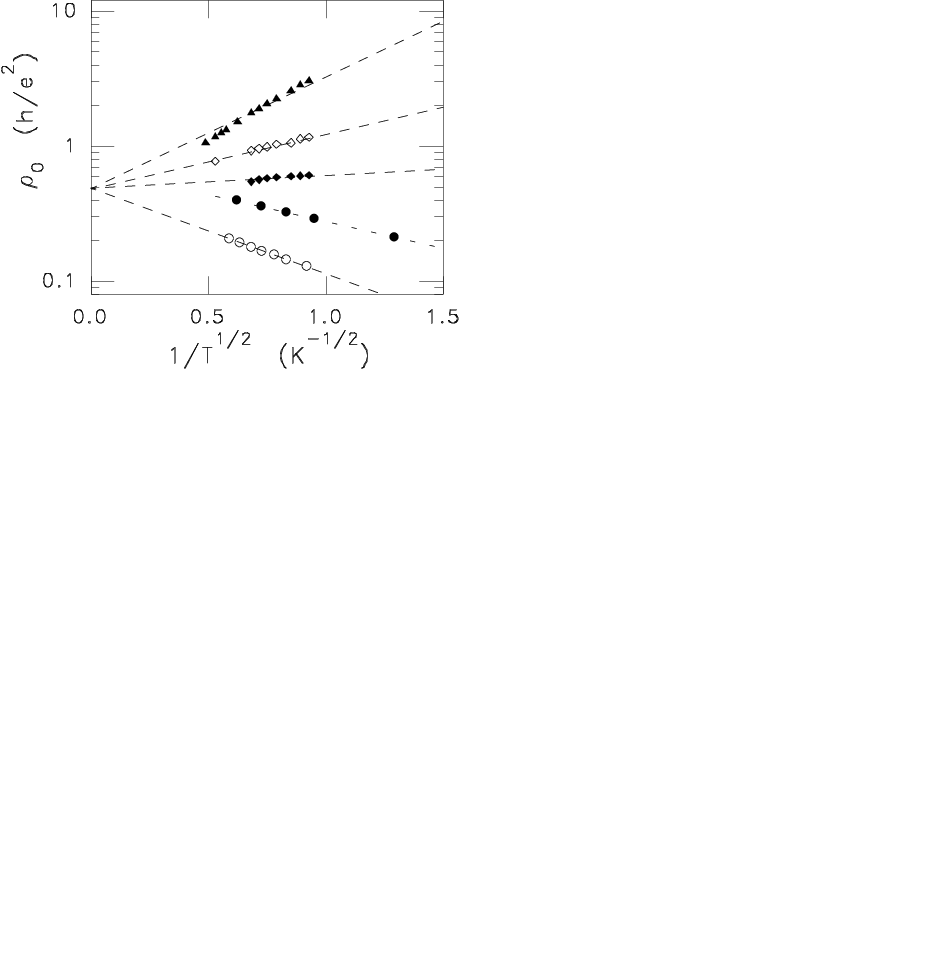}
\caption {
Zero field resistivities for densities (in units of 10$^{15}$m$^{-2}$) 
of .48($\blacktriangle$); .59($\lozenge$); .67 ($\blacklozenge$);
1.2($\bullet$); 1.5 ($\circ$). Lines are only a guide to the eye.
}
\label{fig2}
\end{figure}

       These results demonstrate a metal-insulator
transition, of the same type seen in Si-MOSFETs. This can
be seen more clearly in figure~\ref{fig2} where data over the
range 1-4K, where the metallic behaviour was observed, has
been replotted with extra results obtained using the PPC
effect to change the density. The transition occurs at a
critical density of about 1$\times$10$^{15}$m$^{-2}$ with a critical
resistivity, $\rho_c$, of order 0.5h/e$^2$. On the two sides of the
transition the resistivity varies approximately as
$\rho_c \, \exp [\pm (T_0/T)^{1/2}]$.  At the lowest temperatures this
implies a resistance going to zero, as is observed for
example in superconducting-insulating transitions\cite{r18}, but
here, as in the Si-MOSFETs, the resistivity drop in the
metallic phase in fact eventually saturates at a constant
value\cite{r19}.

       Although the $\exp (a/T^{1/2})$ dependence on the
insulating side of the transition suggests that variable
range hopping, with a Coulomb gap, may play an important
role\cite{r20} this is not necessarily the case. A Coulomb gap
provides no obvious explanation for the enhanced
conductivity on the metallic side of the transition but
the temperature dependence on both sides of the transition
can be naturally explained in terms of scaling behaviour
for an interacting system\cite{r6}. The predicted behaviour,
around the critical density n$_c$ , is given by
\begin{displaymath}
                 \rho _{xx} (\delta n,T) = 
      \rho _c \, \exp(-A \delta n/T ^{1/z \nu}) 
\end{displaymath}
where A is an unknown constant, $\delta n = (n-n_c )/n_c$  and z and
$\nu$ are respectively the dynamical and correlation length
exponents. The value of z$\nu$ then determines the temperature
dependence in both the insulating and metallic phases. 
Data from Si-MOSFETs agrees well with this
expression\cite{r1,r2,r5} using $z\nu \approx 1.6$.

\begin{figure}[tbh]
\vspace{7.2cm}
\includegraphics{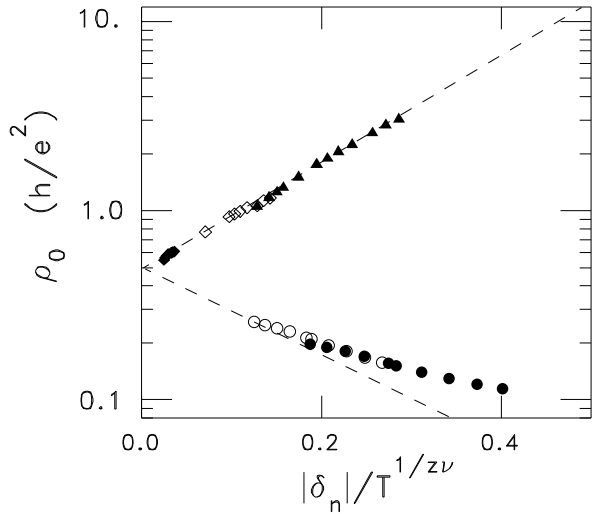}
\caption {
Scaling plot (with $z\nu = 1.6$) for data from two samples with critical
densities (n$_c$) of 1.0 $\times$ 10$^{15}$ m$^{-2}$(sample E) and
 0.7 $\times$ 10$^{15}$ m$^{-2}$ (sample F). The density was varied using the
PPC effect and temperatures measured in kelvins. Dotted lines have slopes of
equal magnitude but opposite sign.
}
\label{fig3}
\end{figure}

       Fig.~\ref{fig3} shows the p--SiGe results plotted in this way for
two samples (E and F) where the density was varied using
the PPC effect. Both samples had a spacer layer of 20nm
but different doping levels.  The general behaviour is
explained well although slightly different values of $n_c$
(.7 and 1.0 $\times$ 10$^{15}$ m$^{-2}$) are required in each case to
collapse the data to a single curve. Also slightly
different values of the coefficient A appear to be
required in each case so full duality is not observed\cite{r21}. This is
not surprising as there is no reason to expect
changes in density produced by changing the doping level
or by exploiting the PPC effect to be equivalent. Because
of the relatively small range of densities the exponent $z\nu$
cannot be determined with any precision from these
measurements but values of  $z\nu = 1.6$, used in fig.~\ref{fig3} for
comparison with the Si-MOSFETs, or $z\nu = 2$ (suggested by
fig.~\ref{fig2}) are both at least approximately correct.

       The scaling theory accounts naturally for the
existence of a metal-insulator transition and the duality
between the temperature dependence in the metallic and
insulating phases. It does not identify any specific
mechanism driving the transition. Strong many-body
interactions, as suggested previously\cite{r22}, are an obvious
possibility; another could be the spin-gap, suggested by
Pudalov\cite{r4}, which occurs because the spin-orbit splitting
that is non-zero even at B = 0.  

       The similarities between the metal-insulator
transition in p-SiGe and Si-MOSFETs suggests the same
mechanism is driving the transition in the two cases. In
both systems many-body effects (electron-electron or hole-
hole interactions) are an order of magnitude larger than
the Fermi energy, in contrast to the situation in more
conventional 2D systems such as GaAs/GaAlAs
heterojunctions, where the two energies are of a very
similar magnitude. In Si-MOSFETs the characteristic and
unconventional temperature dependence in the metallic
phase is seen only in high mobility samples: p-SiGe
samples showing the same effects also have high
mobilities. Spin-orbit effects are probably important in
both cases, in Si-MOSFETs because of the strongly
asymmetric confining potential and in p-SiGe because the
holes are in $|M_J| = 3/2$ states.  With the holes in an
asymmetric potential well, a spin-gap, at B=0 should also
be a possibility in p-SiGe. For the spin-gap to not be
washed out large angle scattering processes must dominate
in producing the resistance. This is the case here, in
both systems, with the quantum and transport mobilities
essentially the same.  Finally, it is also perhaps
noteworthy that in both systems an insulating phase is
also observed around $\nu = 3/2$ \cite{r1,r15,r16}
although for SiGe this is complicated by a ferromagnetic--paramagnetic
phase transition that occurs when the 0$\uparrow$ and 1$\downarrow$
Landau levels cross. It is not clear at this stage which of these
several factors are required for the occurrence of the B = 0
 metal-insulator transition.

       Experimental results are reported showing that low
density p-SiGe samples exhibit the same kind of metal-
insulator transition at B = 0 seen in Si-MOSFETs. In
particular the resistivity in the metallic phase is
unconventional, decreasing exponentially with temperature.
This behaviour confirms other experimental observations
that a metal-insulator transition is allowed in 2D
systems, at B = 0. It is also consistent with predictions
that the metallic behaviour is not Fermi liquid like.
Interaction effects are large in both systems and spin-
orbit effects are also possibly important. Either, or
both, of these properties can reconcile the apparent
conflict between the experimental observation of a metal-
insulator transition and the predictions of one parameter
scaling theory and can account for the anomalous metallic
behaviour.

       We would like to thank S.V. Kravchenko, V.M.
Pudalov and A.S. Sachrajda for useful discussions.


\begin{references}
\bibitem{r19} Measurements at the lowest temperatures were
       complicated, in some cases, by the Al bonding wires
       used to contact the Hall bar which provide the main
       thermal link to the carriers. When this is broken
       by the Al going superconducting very long time
       constants are introduced and it becomes difficult
       to establish the precise temperature of the holes.
\bibitem{r21} Although it should be noted that subtraction of a
       constant impurity resistivity from the experimental
       data could significantly improve this.
\end{references}
\end{document}